\documentclass[twocolumn]{aastex631}

\usepackage{commath}
\usepackage{floatrow}
\usepackage{soul}
\usepackage{CJK}
\usepackage{threeparttable}
\usepackage{soul}
\usepackage{ulem}

\newcommand{\oiii}{\hbox{[O\,{\scriptsize III}]}}
\newcommand{\nii}{\hbox{[N\,{\scriptsize II}]}}
\newcommand{\oii}{\hbox{[O\,{\scriptsize II}]}}
\newcommand{\hii}{\hbox{H\,{\scriptsize II}}}
\newcommand{\hi}{\hbox{H\,{\scriptsize I}}}

\shorttitle{MaNGA 8313-1901}
\shortauthors{Ju et al.}
\graphicspath{{./}{figures/}}
\begin{document}
\begin{CJK*}{UTF8}{gbsn}

\title{ MaNGA 8313-1901: gas accretion observed in a blue compact dwarf galaxy? }

\correspondingauthor{Jun Yin}
\email{jyin@shao.ac.cn}

\author[0000-0002-5815-2387]{Mengting Ju (居梦婷)}
\affiliation{Key Laboratory for Research in Galaxies and Cosmology, Shanghai Astronomical Observatory, Chinese Academy of Sciences, 80 Nandan Road, Shanghai 200030, People's Republic of China}
\affiliation{University of Chinese Academy of Sciences, 19A Yuquan Road, Beijing 100049, People's Republic of China}

\author[0000-0002-4499-1956]{Jun Yin (尹君)}
\affiliation{Key Laboratory for Research in Galaxies and Cosmology, Shanghai Astronomical Observatory, Chinese Academy of Sciences, 80 Nandan Road, Shanghai 200030, People's Republic of China}
\affiliation{Key Lab for Astrophysics, Shanghai, 200034, People's Republic of China}

\author[0000-0001-5968-1144]{Rongrong Liu (刘蓉蓉)}
\affiliation{Key Laboratory for Research in Galaxies and Cosmology, Shanghai Astronomical Observatory, Chinese Academy of Sciences, 80 Nandan Road, Shanghai 200030, People's Republic of China}
\affiliation{University of Chinese Academy of Sciences, 19A Yuquan Road, Beijing 100049, People's Republic of China}

\author[0000-0003-2478-9723]{Lei Hao (郝蕾)}
\affiliation{Key Laboratory for Research in Galaxies and Cosmology, Shanghai Astronomical Observatory, Chinese Academy of Sciences, 80 Nandan Road, Shanghai 200030, People's Republic of China}

\author[0000-0001-8611-2465]{Zhengyi Shao (邵正义)}
\affiliation{Key Laboratory for Research in Galaxies and Cosmology, Shanghai Astronomical Observatory, Chinese Academy of Sciences, 80 Nandan Road, Shanghai 200030, People's Republic of China}
\affiliation{Key Lab for Astrophysics, Shanghai, 200034, People's Republic of China}

\author[0000-0002-9767-9237]{Shuai Feng (冯帅)}
\affiliation{College of Physics, Hebei Normal University, 20 South Erhuan Road, Shijiazhuang 050024, People's Republic of China}
\affiliation{Hebei Key Laboratory of Photophysics Research and Application, Shijiazhuang 050024, People's Republic of China}

\author[0000-0002-1321-1320]{Rog\'erio Riffel}
\affiliation{Departamento de Astronomia, Instituto de F\'\i sica, Universidade Federal do Rio Grande do Sul, CP 15051, 91501-970, Porto Alegre, RS, Brazil}
\affiliation{Laborat\'orio Interinstitucional de e-Astronomia - LIneA, Rua Gal. Jos\'e Cristino 77, Rio de Janeiro, RJ - 20921-400, Brazil}

\author[0000-0001-5561-2010]{Chenxu Liu (刘辰旭)}
\affiliation{Department of Astronomy, The University of Texas at Austin, 2515 Speedway Boulevard, Austin, TX 78712, USA}

\author[0000-0002-3746-2853]{David V. Stark}
\affiliation{Departments of Physics and Astronomy, Haverford College, 370 Lancaster Ave, Haverford, PA 19041, USA}
\affiliation{Department of Astronomy, University of Washington, 3910 15th Ave. NE, Room C319, Seattle, WA 98195-0002, USA}

\author[0000-0002-3073-5871]{Shiyin Shen (沈世银)}
\affiliation{Key Laboratory for Research in Galaxies and Cosmology, Shanghai Astronomical Observatory, Chinese Academy of Sciences, 80 Nandan Road, Shanghai 200030, People's Republic of China}
\affiliation{Key Lab for Astrophysics, Shanghai, 200034, People's Republic of China}

\author{Eduardo Telles}
\affiliation{Observat\'orio Nacional R. Gen. Jos\'e Cristino, 77, 20921-400, Rio de Janeiro, Brazil}

\author[0000-0003-3526-5052]{Jos\'e G. Fern\'andez-Trincado}
\affiliation{Instituto de Astronom\'ia, Universidad Cat\'olica del Norte, Av. Angamos 0610, Antofagasta, Chile}

\author[0000-0003-4874-0369]{Junfeng Wang (王俊峰)}
\affiliation{Department of Astronomy, Physics Building, Xiamen University, Xiamen, Fujian, 361005, Peopleʼs Republic of China} 

\author{Haiguang Xu (徐海光)}
\affiliation{School of Physics and Astronomy, Shanghai Jiao Tong University, 800 Dongchuan Road, Shanghai 200240, People's Republic of China}

\author[0000-0002-3601-133X]{Dmitry Bizyaev}
\affiliation{Apache Point Observatory and New Mexico State University, P.O. Box 59, Sunspot, NM, 88349-0059, USA}
\affiliation{Sternberg Astronomical Institute, Moscow State University, Moscow, Russia}

\author[0000-0002-2204-6558]{Yu Rong (容昱)}
\affiliation{Department of Astronomy, University of Science and Technology of China, No.96, JinZhai Road, Baohe District, Hefei, Anhui, 230026, People's Republic of China}

\begin{abstract}

Gas accretion is an important process in the evolution of galaxies, but it has limited direct observational evidences. In this paper, we report the detection of a possible ongoing gas accretion event in a Blue Compact Dwarf (BCD) galaxy, MaNGA 8313-1901, observed by the Mapping Nearby Galaxies and Apache Point Observatory (MaNGA) program. This galaxy has a distinct off-centered blue clump to the northeast (the NE clump) that shows low metallicity and enhanced star-formation. The kinematics of the gas in the NE clump also seems to be detached from the host BCD galaxy. Together with the metallicity drop of the NE clump, it suggests that the NE clump likely has an external origin, such as the gas accretion or galaxy interaction, rather than an internal origin, such as an \hii~complex in the disk. After removing the underlying host component, we find that the spectrum of the ``pure" clump can match very well with a modeled spectrum containing a stellar population of the young stars ($\le 7$ Myr) only. This may imply that the galaxy is experiencing an accretion of cold gas, instead of a merger event involving galaxies with significant pre-existing old stars. We also find signs of another clump (the SW clump) at the south-west corner of the host galaxy, and the two clumps may share the same origin of gas accretion.

\end{abstract}


\section{Introduction} \label{sec:intro}

Numerical simulations suggest that the accretion of the metal-poor gas from the cosmic web is a common process to fuel the formation of disk galaxies \citep{Finlator2008,Stewart2011,Altay2013}. Gas accretion, often with lower metallicity compared to surroundings, might dominate the early growth of galaxies \citep{Finlator2008}. 
\cite{kennicutt1998} shows there is a tight relation between the star formation rates (SFRs) and the total mass of gas, i.e., the Kennicutt-Schmidt law. However, the depletion time, defined as the gas mass divided by the SFR observed at the present day, is shorter than the cosmic time in many star-forming galaxies. This implies that there should be gas replenishment in their evolutionary history.
For very nearby galaxies, such as the Milky Way, M31, and M33, the pristine gas accretion is also needed to explain the observed metallicity distribution function and the metallicity radial gradient \citep{Yin2009,kang2012}.

The characteristics of the gas accretion seems to be related with the dark matter masses of the host halos where the galaxies are located in \citep{Dekel2009}. The most massive halos ($\text{M}_\text{halo} > 10^{14}\ \text{M}_{\sun}$) are usually well virialized in the local universe. The radiation pressure competes with the gravity, making the cold gas hard to fall in. Gas accretion can happen more easily in less massive halos, where the cold gas can directly feed the star formation of the galaxies \citep{Fumagalli2011,genel2012,Liu2019}. 

However, it is difficult to directly observe the accreting gas, as they are often diffused and extend to very large scales ($\sim$ 100 kpc), making the surface densities fall below the detection limit of most large surveys. There are many indirect observational evidences of gas accretion \citep{Cresci2010, almeida2014review}. For example, the metallicity inhomogeneities, particularly the combination phenomena that the metallicity drops accompanied with intense starbursts, as observed in some Blue Compact Dwarf (BCD) galaxies, could be one evidence of cold-flow accretion \citep{Crain2009,Almeida2015,mannucci2010,Lassen2021}.
There are simulations that can reproduce this phenomena as well, in that the external gas accreted by host galaxies can lower the metallicity of star-forming gas and trigger the star formation simultaneously \citep{Yates2012}. 

The strong star-forming activities triggered by the gas accretion can produce bright off-centered starburst clumps.
This kind of off-centered starburst clumps are very abundant in high redshift galaxies \citep{mannucci2010,Rauch2011,Elmegreen2013}. As good analogs of these high redshift star-forming galaxies \citep{Papaderos2012}, local dwarf galaxies also show various signs of star formation in their outskirts. \cite{Cignoni2019} discussed the tendency for dwarf irregulars to host young populations at large radii. Some nearby galaxies show extended UV disks, which means there may be star formation in their outer disks \citep{Thilker2007}. In addition, high velocity clouds (HVCs) are found in many nearby galaxies including our own Milky Way \citep{Hernandez2013,Fraternali2014}, which may provide a source of material to be accreted to sustain the star formation in these galaxies. These local galaxies can give us suggestive clues to study the gas accretion event. 

The large Integral Field Unit (IFU) surveys of nearby galaxies, such as the Mapping Nearby Galaxies and Apache Point Observatory survey \citep[MaNGA;][]{bundy2015}, provide various spatially-resolved information that can be critical in studying possible gas accretion processes.
For example, \cite{Hwang2019} analyzed the data from the MaNGA survey and found that some off-centered anomalously low-metallicity (ALM) regions, defined as having metallicity more than $0.14$ dex lower than that implied by the stellar mass density and metallicity relationship ($\Sigma_*-Z$), may be produced by the impulsive accretion of gas. These ALM regions have higher specific SFRs and younger stellar populations, and they are more easily to be found in the blue low-mass galaxies than other galaxies.  Employing a simple but intuitive inflow model, \cite{pace2021} found that the metallicity profiles of the low-mass and non-interacting galaxies with ALM regions can be produced by large inflows. Some of the H$\alpha$ blobs \citep{Lin2017,Pan2020,ji2021} may also be associated with gas accretion. The IFU observations can also provide spatially-resolved kinematic information, which sometimes is crucial to identifying unique patterns induced by the gas inflows \citep{chen2016,jin2016}.

In this paper, we present a BCD galaxy, MaNGA 8313-1901, with a bright clump that may indicate a gas accretion scenario. 
The paper is organized as follows: In Section~\ref{sec:data}, we introduce the Dark Energy Spectroscopic Instrument (DESI) Legacy Imaging data and the MaNGA survey data of MaNGA 8313-1901. In Section~\ref{sec:method}, we analyze the structural and spectral properties of this galaxy. In Section~\ref{sec:discussion}, we discuss the possible gas accretion scenario for this galaxy. We summarize this paper in Section~\ref{summary}. Throughout this paper, we adopt an the cosmological parameters of $H_0=70\ \rm km/s/Mpc, \Omega_{M} = 0.3$, and $\Omega_\Lambda=0.7$.


\begin{table}
\caption{Properties of MaNGA 8313-1901}

\begin{tabular}{@{}lrr@{}}
\hline
\multicolumn{3}{c}{MaNGA 8313-1901 $=$ SDSS\,J160108.90+415250.7} \\
\hline
Parameters & \multicolumn{2}{c}{Data} \\ 
\hline
$\text{MaNGA ID}$              & \multicolumn{2}{c}{1-248352} \\
$\mathrm{RA}$ (J$2000$)          & \multicolumn{2}{c}{16:01:08.90}\\
                                 &\multicolumn{2}{c}{(240.28712$^\circ$)}\\ 
$\mathrm{DEC}$ (J$2000$)         & \multicolumn{2}{c}{+41:52:50.77}\\
                                 &\multicolumn{2}{c}{(41.88075$^\circ$)}\\ 
$z$\,$^{a}$                              & \multicolumn{2}{c}{$0.02425$}\\ 
$d$ [Mpc]                        & \multicolumn{2}{c}{$103.932$}\\ 
\noalign{\smallskip}
$M_{\rm NUV}$~[mag]\,$^{a}$                    & \multicolumn{2}{c}{-17.46}\\
$M_g$~[mag]\,$^{a}$                            & \multicolumn{2}{c}{-18.64}\\
$M_r$~[mag]\,$^{a}$                            & \multicolumn{2}{c}{-18.86}\\
$M_z$~[mag]\,$^{a}$                            & \multicolumn{2}{c}{-19.03}\\
\noalign{\smallskip}
\hline
log(M$_{*}$/M$_{\odot}$)\,$^{a}$         & \multicolumn{2}{c}{$8.88$}\\       
log(M$_{*}$/M$_{\odot}$)\,$^{b}$         & \multicolumn{2}{c}{$9.28$}\\  
log(M$_\text{HI}$/M$_{\odot}$)\,$^{c}$    & \multicolumn{2}{c}{$9.37$}\\ 
log(M$_\text{halo}$/M$_{\odot}$)\,$^{d}$  & \multicolumn{2}{c}{$11.03$}\\

\noalign{\smallskip}
\hline
sersic index \,$^{e}$                  & \multicolumn{2}{c}{1.46 $\pm$ 0.04 (the host galaxy)}\\
                               & \multicolumn{2}{c}{0.19 $\pm$ 0.04 (the NE clump)}\\
                               
effective radius\,$^{e}$ [kpc]         & \multicolumn{2}{c}{1.23 $\pm$ 0.13 (the host galaxy)}\\
                               & \multicolumn{2}{c}{0.29 $\pm$ 0.01 (the NE clump)}\\

\noalign{\smallskip}
\hline
\end{tabular}\\
\scriptsize{
$^{a}$ The NASA-Sloan Atlas catalog: http://www.nsatlas.org\\
$^{b}$ The MPA - JHU catalog \citep{Kauffmann2003b}\\
$^{c}$ \cite{HI2019}\\
$^{d}$ \cite{Yang2007,Yang2012}\\
$^{e}$ See more details in Section~\ref{sec:method}}
\label{tab:table}      
\end{table}

\section{DATA} \label{sec:data}

In this work, we use the photometric data from the DESI Legacy Imaging Surveys \citep{DESI2019} and the spatially resolved spectroscopic data from the MaNGA survey \citep[MaNGA;][]{bundy2015}.
MaNGA 8313-1901 was imaged by the DESI Legacy Imaging Surveys \citep{DESI2019}. The Legacy surveys provide images of three optical bands ($g, r$ and $z$) covering about 14,000 deg$^2$. They include three public projects: The Beijing-Arizona Sky Survey \citep[BASS;][]{BASS2017} takes images at $\rm32^\circ \le DEC\le 84^\circ$ in $g$-band and $r$-band; The Mayall $z$-band Legacy Survey \citep[MzLS;][]{mzls2016} provides the $z$-band images with the same region as BASS; The Dark Energy Camera Legacy Survey \citep[DECaLS;][]{DECALS2016} targets the remaining $\approx9350$ deg$^2$ with $g,r$ and $z$-bands. In this paper, we only use the coadded $g$-band image from BASS (the Legacy Surveys DR8) to analyze the morphological structure of our target (Section \ref{sec:DESI}). BASS uses the 90Prime camera \citep{Williams2004} on the Bok 2.3-m telescope. The median 5$\sigma$ detection limits of BASS are $g$ = 23.48 and $r$ = 22.87 AB mag which are about 1~mag deeper than those of the SDSS images.

The MaNGA survey has observed approximately 10,000 nearby galaxies within the redshift range of $0.01<z<0.05$ as part of the fourth generation of the Sloan Digital Sky Survey \citep[SDSS-IV;][]{blanton2017}. The survey uses 17 different hexagonal fiber-bundles IFUs ranging from 19 fibers spanning a field of view (FoV) of 12\arcsec\ in diameter to 127 fibers covering a FoV of 32\arcsec\ in diameter \citep{drory2015}. All fibres are fed into the Baryon Oscillation Spectroscopic Survey (BOSS) spectrographs \citep{smee13} on the SDSS 2.5-meter telescope at Apache Point Observatory \citep{gunn06}. The site seeing is $\sim$ 1.5\arcsec, and the point spread function (PSF) in a typical MaNGA reconstructed data cube can be characterized by a circular Gaussian distribution with a full width at half-maximum (FWHM) of 2.5\arcsec\ \citep{Law2015,yan2016b}. The MaNGA spectrograph provides wavelength coverage from 3600\,\AA\ to 10,300\,\AA~with a spectral resolution of $R \sim 1100-2200$ \citep{law2016,Law2021,yan2016a}. The MaNGA data used in this work is taken from the Data Analysis Pipeline (DAP) hybrid binning (HYB) MPL-8 version (see \citealp{westfall2019} and \citealp{Belfiore2019} for more details). The data cube has a spatial grid of 0.5\arcsec$\times$ 0.5\arcsec.

MaNGA 8313-1901 is one of the BCDs in the MaNGA survey. It is the most noticeable galaxy in the BCD sample as having a large off-centered clump. MaNGA 8313-1901 is a low-redshift galaxy at $z = 0.02425$ with a stellar mass of log($\text{M}_*/\text{M}_{\odot}$) = 8.88 (the NASA-Sloan Atlas catalog, although the stellar mass is also estimated to be log($\text{M}_*/\text{M}_{\odot}$) = 9.28 by the Max Planck for Astrophysics (MPA) - Johns Hopkins University (JHU) catalog \citep{Kauffmann2003b}). The main properties of this galaxy are listed in Table~\ref{tab:table}.
MaNGA 8313-1901 shows a high average surface brightness of $\mu_{g,r_{50}} = 19.89\ \rm mag/arcsec^{2}$ within the half light radius $r_{50}$ \citep{mangasample}. In the Galaxy Zoo 2 classifications, this galaxy is unbarred \citep{2013MNRAS.435.2835W}.

\begin{figure*}
    \centering
    \includegraphics[width=1.0\textwidth,clip,trim={0 0 0 0}]{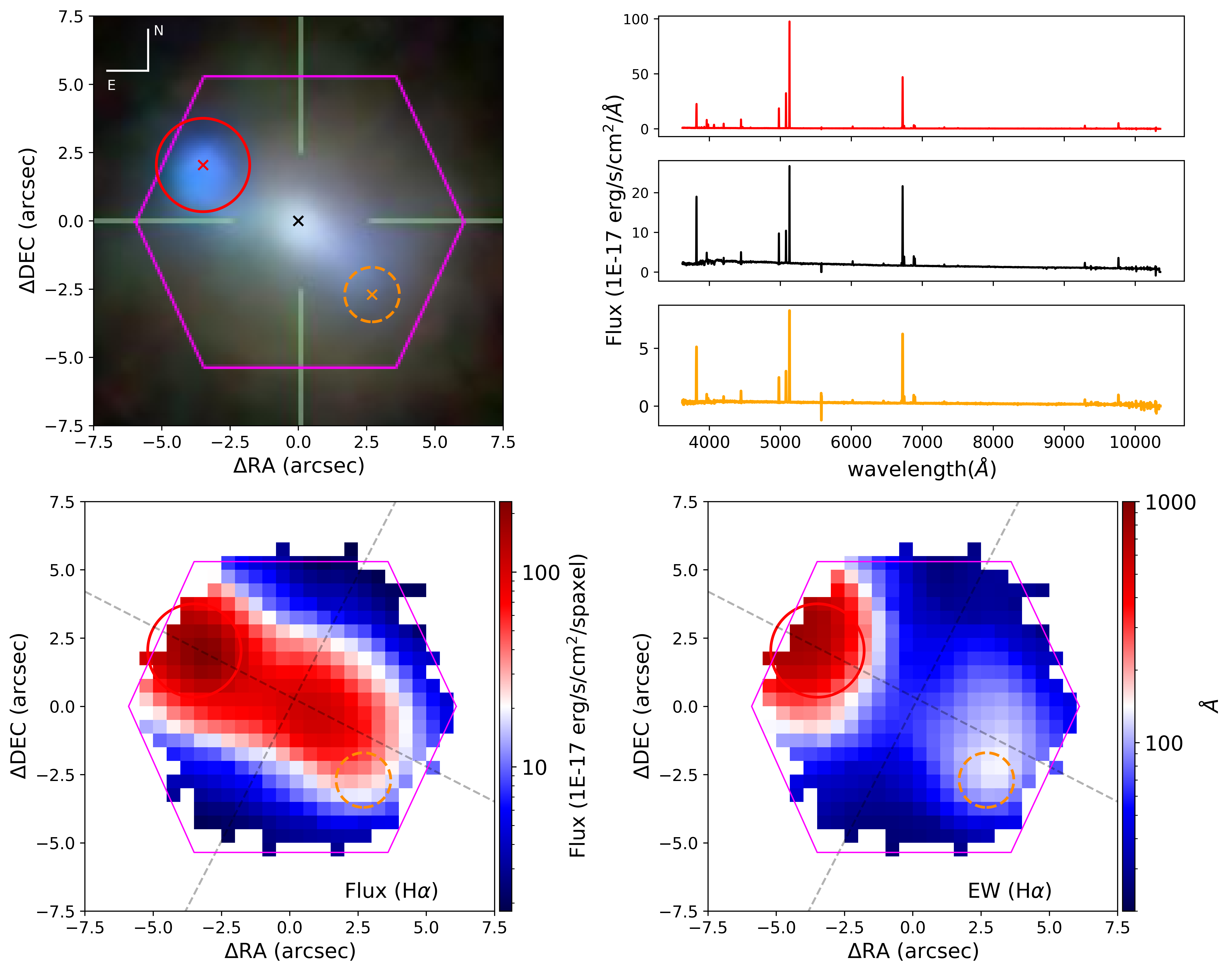} 
    \caption{The top left panel shows the SDSS $gri$ composite image, which covers 15\arcsec $\times$ 15\arcsec\ in size and corresponds to 7.56 kpc $\times$7.56 kpc at its distance. The magenta hexagon shows the coverage of the MaNGA bundle in this field. The red solid circle and the orange dashed circle are the locations of the NE clump and the SW clump, defined in Section \ref{sec:DESI} and \ref{sec:sfr}, respectively. In the top right panel, the spectra of three spexals indicated as red, black, and orange crosses in the top left panel are shown as the red, black, and orange solid lines, respectively. Bottom left: the dust-corrected flux map of the H$\alpha$ emission line. Bottom right: the Equivalent Width map of the H$\alpha$ emission line. The major and minor axes are shown as dashed grey lines. The position angle is adopted from the NASA-Sloan Atlas catalog.}
    \label{fig:image}
\end{figure*}

In the top left panel of Fig.~\ref{fig:image}, we show the SDSS {\it gri}-band composite image. 
It covers 15\arcsec $\times$ 15\arcsec, corresponding to 7.56 kpc $\times$ 7.56 kpc at its redshift. The magenta hexagon indicates the coverage of the 19-fiber bundle of the MaNGA survey. The FWHM of PSF of this galaxy is 2.59\arcsec\ (1.31 kpc) in $g$-band. In this image, there is a giant clump that is located to the north-east direction of the host galaxy, which we name as the NE clump.  

\label{emission line}
In the top right panel of Fig.~\ref{fig:image}, we show the spectra at the three spaxels in the galaxy, labeled as the red, black and orange crosses in the top left panel, respectively. The red cross is at the center of the NE clump, which will be more specifically defined in Section~\ref{sec:DESI}. The black cross is at the galaxy center. The orange cross is at the center of the other clump, which will be defined in Section~\ref{sec:sfr}. The three spectra show strong emission lines.
The emission lines of the NE clump at the center (the red spectrum) are significantly stronger than those of the center of the host galaxy (the black spectrum). 

We select 376 spaxels ($\sim$ 80\%) with signal-to-noise ratio (S/N) of the H$\alpha$ emission line higher than 5 from the MaNGA DAP data for further analysis.
These spaxels all fall into the star formation region in the BPT diagram \citep{BPT1981,kewley2001,kauffmann2003a}.
In the bottom left panel of Fig.~\ref{fig:image}, the flux map of the H$\alpha$ emission line shows two strong peaks. The stronger H$\alpha$ peak is located in the NE clump, while the other one overlaps well with the center of the host galaxy. The H$\alpha$ flux is corrected for the dust attenuation using the H$\alpha$/H$\beta$ ratio. The intrinsic line ratio is assumed to be 2.86 and the attenuation curve is adopted as in \cite{Calzetti2000}. The spectra that we use to analyze the stellar population are corrected for the Milky-Way extinction law \citep{Fitzpatrick1999}. The bottom right panel demonstrates the equivalent width (EW) map of the H$\alpha$ emission line. The EW(H$\alpha$) in the NE clump is dramatically higher than that of all other spaxels of the galaxy. 

\section{The physical properties of MaNGA 8313-1901}\label{sec:method}

\subsection{The optical morphology}\label{sec:DESI}

\begin{figure*}
    \centering
    \includegraphics[width=1\textwidth,trim={0 0 0 0},clip]{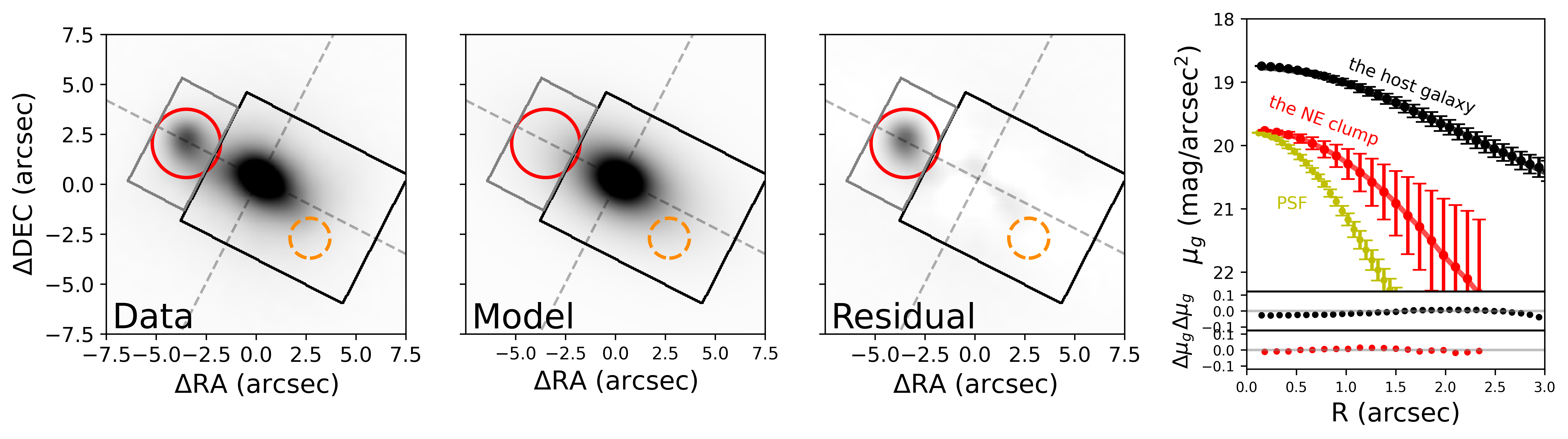}
    \caption{Two-dimensional surface brightness modeling of the host galaxy of MaNGA 8313-1901 with {\tt GALFIT}. The leftmost panel shows the observed $g$-band image of BASS with a pixel size of 0.06\arcsec\ and a PSF FWHM of 1.68\arcsec. The black box is defined to be the fitting area of the host galaxy. The grey box with a size of $\sim$ 3\arcsec $\times$ 6\arcsec and centered at the peak of H$\alpha$ flux in the NE clump is the region we fit to obtain the morphological parameters of the NE clump with a size of $\sim$ 3\arcsec $\times$ 6\arcsec. The best-fitted 2D model of the host galaxy given by {\tt GALFIT} and the host subtracted residual image are shown in the second and third panels. The color scale is the same for all images, and ranges from $-0.25$ to 20 nanomaggies/arcsec$^2$. The two grey dashed lines show the major (northeast-southwest direction) and the minor (northwest-southeast direction) axes. In the left three panels, the red solid circle indicates our definition of the NE clump region with a radius of 1.71\arcsec. The orange dashed circle with a radius of 1\arcsec\ shows our definition of the SW clump (Section~\ref{sec:sfr}). The rightmost panel shows the surface brightness profiles of the host galaxy (black), the NE clump (red), and the scaled PSF model (yellow). The dots with error bars demonstrate the observed data, and the lines stand for the 1D profiles of the best-fit {\tt GALFIT} model. The two bottom sub-panels of the rightmost panel are the residual ($\mu_{\text{data}}-\mu_{\text{model}}$) profiles of the surface brightness of the host galaxy (black) and the NE clump (red) respectively.}
    \label{fig:galfit}
\end{figure*}

Both the SDSS image (the top left panel of Fig.~\ref{fig:image}) and the Legacy $g$-band image (the leftmost panel of Fig.~\ref{fig:galfit}) show that the NE clump is probably structurally detached from the host galaxy. Therefore, we would like to separate the host galaxy and the NE clump morphologically. 

We use the $\tt GALFIT~v3.0.5$ \citep{peng2002,peng2010}, a two-dimensional fitting methodology software, to do the decomposition. 
According to the axis ratio and the position angle (PA) values listed in the NASA-Sloan Atlas catalog, we draw the major and minor axes with grey dashed lines in Fig.~\ref{fig:galfit}. We define a fitting area shown as the black box, whose sides are parallel to the major and minor axes. The box is drawn to try to have little contamination from the NE clump, yet also to be as big as possible for characterizing the host.
We fit the host galaxy within the black box by convolving a single S\'ersic model with a PSF extracted from an unsaturated star with a high S/N (ID 391 in the eighth Data Release, DR8, of DESI) near MaNGA 8313-1901. When fitting the host galaxy we fix the PA to be the value taken from the NASA-Sloan Atlas catalog. Changing the PA value a little bit will not affect the fitting results much.
From left to right, Fig.~\ref{fig:galfit} exhibits the $g$-band image taken from the DESI Legacy Imageing Survey, the best-fit 2D model of the host galaxy, the host-subtracted residual image, and the 1D surface brightness profile. The color scale is the same for all three images, and ranges from $-0.25$ to 20 nanomaggies\footnote{The definition of the unit ``nanomaggy" can be found in https://www.sdss.org/dr17/algorithms/magnitudes/}/arcsec$^{2}$. The NE clump is clearly shown in the residual image. The best-fit S\'ersic index and the effective radius of the host component are $n=1.46\pm0.04$ and $r_e=1.23\pm0.13$ kpc ($2.44\arcsec\pm0.26\arcsec$). These two parameters are comparable to typical values of BCDs in general \citep{Amorin2009}.

We also use $\tt GALFIT$ to analyze the structure of the NE clump using the residual image after subtracting the best-fit host galaxy, as shown in the right panel of Fig.~\ref{fig:galfit}. We use a single S\'ersic model and the same PSF model extracted from ID 391 to fit the image in the grey box ($\sim$ 3\arcsec\ $\times$ 6\arcsec) defined to include mostly the NE clump.
The best-fit S\'ersic index and the effective radius of the fitted NE clump are $n=0.19\pm0.04$ and $r_e=290\pm12$ pc ($0.57\arcsec\pm0.24\arcsec$). The radial profiles of the host galaxy (black dots) and the NE clump (red dots) are shown in the rightmost panel of Fig.~\ref{fig:galfit}. In this panel, we also show the PSF profile with yellow dots. We can see that the NE clump is more extended than the PSF. Therefore, the size we measured for the NE clump is real. The diameter of the NE clump is 580 pc, which is significantly larger than those of local clumps, and is comparable to the typical sizes ($\sim$ 1 kpc) of the clumps at high redshifts \citep{Lagos2007,Elmegreen2013,Wuyts2014,meng2020}.

We draw a red circle with a radius of 1.71\arcsec\ (3 times of the effective radius of the NE clump) in Fig.~\ref{fig:galfit}, centered at the center of the NE clump obtained from the {\tt GALFIT} fitting. We define this region as the NE clump region of our BCD galaxy.

\subsection{Star formation and Gas-phase metallicity}\label{sec:sfr}

\begin{figure*}
    \includegraphics[width=1\textwidth]{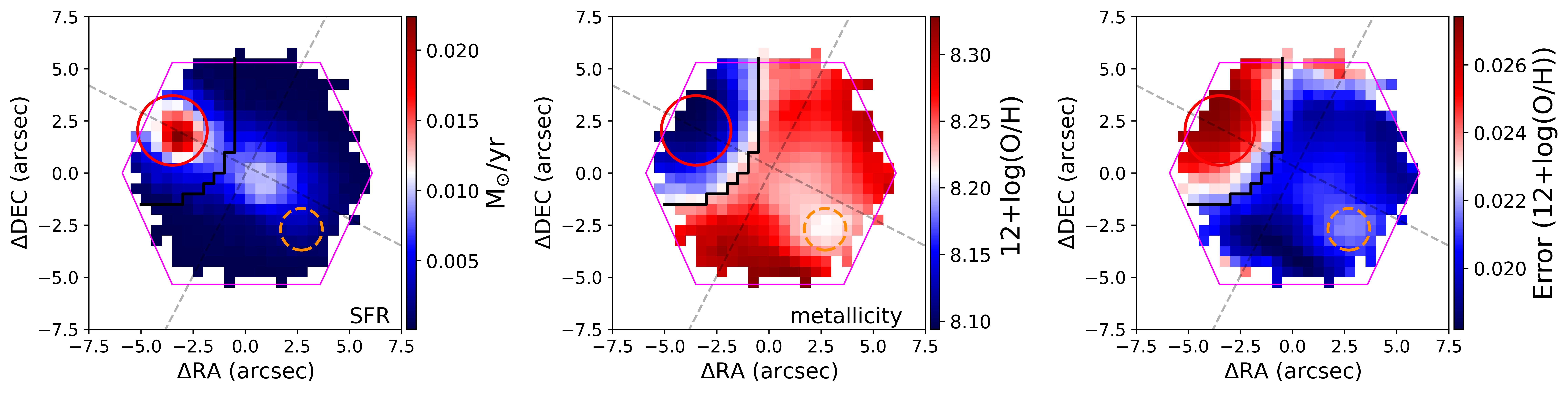}
    \caption{The SFR map (the left panel) and the metallicity map (the middle panel) with the metallicities measured via the O3N2 method. The error map of the metallicity measurements is shown in the right panel. The red circle indicates the NE clump region, and the orange dashed circle shows the SW clump region.}
    \label{fig:maps}
\end{figure*}

With the MaNGA data, we can measure the SFR and the gas-phase metallicity of each spaxel in the data cube, then obtain the map of the SFR and the metallicity for the whole galaxy. Since all of the analyzed spaxels are located in the star-formation branch in the BPT diagram \citep{BPT1981,kewley2001,kauffmann2003a}, we use the dust-corrected  H$\alpha$ luminosity to calibrate the SFR with Equation~\eqref{eq:1}， taken from \cite{kennicutt2012} and \cite{hao2011}. 

\begin{small}
\begin{equation}
    \log\left(\frac{\text{SFR}}{\text{M}_{\odot}/\text{yr}}\right) = \log\left(\frac{L_{\text{H}\alpha}}{\text{erg/s}}\right)-41.27
    \label{eq:1}
\end{equation}
\end{small}

The SFR map is shown in the left panel of Fig.~\ref{fig:maps}. This map shows an enhanced star formation in the NE clump and the value of the SFR in these spaxels is almost twice the SFRs in the spaxels at the galaxy center.
The total SFR of the NE clump region (within the red circle) and that of the rest area, by summing up the SFRs in corresponding spaxels, are $0.417\pm0.013\ \text{M}_{\odot}/\text{yr}$ and $0.829\pm0.017\ \text{M}_{\odot}/\text{yr}$, respectively.

We use the MaNGA data cube and the {\tt MEGACUBE} software to obtain the spatially-resolved stellar populations \citep{Mallmann2018,Riffel2021} of our galaxy. {\tt MEGACUBE} is based on the {\tt STARLIGHT} code \citep{starlight}, but can be applied to a cube data. The stellar mass and age of each spaxel can be obtained by this code. The mass of the NE clump and that of the rest area are $3.33\times10^{8}\text{M}_{\odot}$  and $3.72\times10^{9} \text{M}_{\odot}$, respectively. So the specific SFR of the NE clump region and that of the rest area are $1.24\times10^{-9}$ yr$^{-1}$ and $2.23\times10^{-10}$ yr$^{-1}$.

The significantly-enhanced star formation activity in the NE clump suggests that this region contains massive star clusters that have been recently formed. The integrated flux of H$\beta$ is log L(H$\beta$) = 40.32 [erg/s] which is large. Given its high luminosity and large size, the NE clump is likely to be an ensemble of unresolved massive star clusters \citep{Lagos2011,Telles2018}, rather than one massive cluster, even though it is hard to tell with the current available images of limited spatial resolutions.

\begin{figure}
    \centering
    \includegraphics[width=1.\textwidth,trim=0 0 0 0]{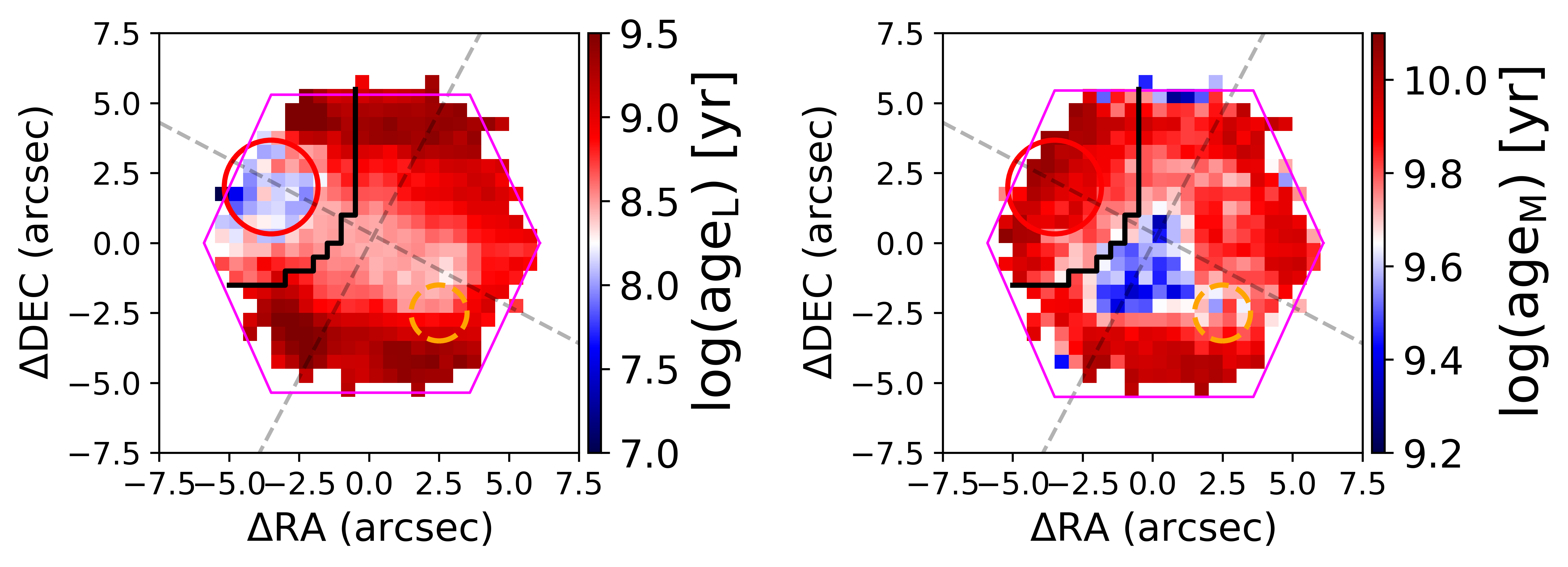}
    \caption{The maps of the light-weighted stellar age (the left panel) and the mass-weighted stellar age (the right panel).}
    \label{fig:meanage}
\end{figure}

In Fig.~\ref{fig:meanage}, we show the light-weighted age map (the left panel) and the mass-weighted age map (the right panel) also obtained from the {\tt MEGACUBE} software. The young stellar population contributes a significantly higher fraction to the light than to the mass. Therefore, the light-weighted stellar age will be severely younger than the mass-weighted age if the young stars dominate the stellar population. This is clearly shown as the difference between these two age maps in the NE clump region in Fig.~\ref{fig:meanage}, indicating that in the NE clump, the young stellar population dominates. 

There are various methods to measure the gas-phase metallicity from the spectrum, such as the direct method and the strong-line methods. The direct method is sometimes referred as the electron temperature ($T_e$) method. 
$T_e$ can be estimated by the strength ratio of an auroral line to a line of lower excitation level of the same element. Usually, $\oiii\lambda4363/\lambda5007$ is used to estimate the $T_e$ in the classical \hii~region model. Unfortunately, $\oiii\lambda4363$ is often too weak to be measurable in normal galaxies.

To overcome this issue, people developed empirical methods, which use various strong emission lines that can be observed in the optical. The popular strong-line methods include the R23 method, where $\rm R23 = \log\frac{\oii\lambda\lambda3727,9 + \oiii\lambda\lambda4959,5007}{H\beta}$ \citep{pagel1979, tremonti2004, henry2013}, the O3N2 method, where $\rm O3N2 = \log\frac{\oiii\lambda5007/H\beta}{\nii\lambda6584/H\alpha}$ \citep{pp04, marino2014}, and the N2 method, where $\rm N2 = \log\frac{\nii\lambda6584}{H\alpha}$ \citep{Storchi1994, Denicolo2002, marino2014}.
\cite{marino2014} calibrated the O3N2 method with observations of \hii~regions based on the direct $T_e$ method:

\begin{small}
\begin{equation}
   12+\mathrm{log(O/H)}=8.533[ \pm0.012 ] - 0.214[\pm0.012] \times \mathrm{O3N2}
   \label{eq:2}
\end{equation} 
\end{small}

This new calibration has been used to measure the metallicity of star-forming galaxies in several works \citep[e.g.][]{Lima2020,ji2021}. In our work, we also use this calibration to measure the metallicity of MaNGA 8313-1901.

In the middle and right panel of Fig.~\ref{fig:maps}, we show the gas-phase metallicity map derived from Equation~(\ref{eq:2}) and its associated map of errors. A significant drop can be seen in the NE clump region. 
We further find another region to the south-west direction of the galaxy also with slightly lower metallicity compared to the host, although the drop of metallicity compared to the host center is within the errorbars. We name this region as the SouthWest (SW) clump, marked by an orange dashed circle with a radius of 1\arcsec\ in Fig.~\ref{fig:maps}. We also label this same circle in previous figures. The SW clump will be discussed in more detail in Section~\ref{sec:discussion}.

\begin{figure*}
    \centering
    \includegraphics[width=1\textwidth,trim={0 0 0 0},clip]{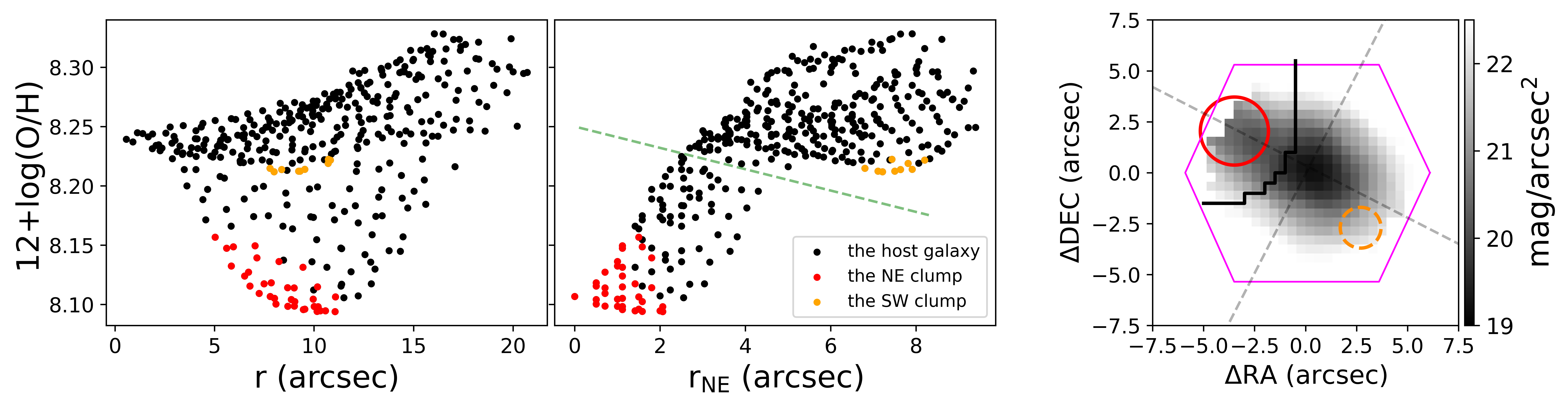}
    \caption{Left: The radial distribution of metallicity for every MaNGA spaxel. Middle: The offset metallicity radial distribution. The radius is  defined as the distance to the spaxel with the highest H$\alpha$ flux in the NE clump. There are two branches in this panel, well separated by the green dashed line. Right: The $g$-band image from the MaNGA survey. The red circle is the NE clump region, and the orange dashed circle shows the SW clump region. The black line is a dividing line of spaxels located between the two branches shown in the middle panel. The red, orange and black data points in the left panel and the middle panel represent the spaxels in the NE clump, the spaxels in the SW clump, and all other remaining spaxels in the right panel, respectively.}
    \label{fig:mse}
\end{figure*}

We show the metallicity of every star-forming spaxel of MaNGA 8313-1901 as a function of galactocentric distance in the left panel of Fig.~\ref{fig:mse}. There is a prominent population of spaxels with low metallicity.
In order to check if these low metallicity spaxels are all associated with the NE clump, we present the metallicity distribution as a function of the distance to the spaxel with the highest H$\alpha$ flux in the NE clump ($\rm r_{NE}$) in the middle panel of Fig.~\ref{fig:mse}. The spaxels in the NE clump (within the red circle in the rightmost panel of Fig.~\ref{fig:mse}) and those in the SW clump (within the orange dashed circle in the rightmost panel of Fig.~\ref{fig:mse}) are colored with red and orange.

In the middle panel of Fig.~\ref{fig:mse}, we find that there are two distinct populations that can be well separated by the green dashed line.
The spaxels with metallicity below the green dashed line are all spatially located in the region northeast of the black line in the $g$-band image integrated from the MaNGA data cube (the right panel of Fig.~\ref{fig:mse}); the spaxels with metallicity above the green dashed line are located in the remaining area.
The low metallicity region, as is shown by the region to the northeast to the black solid line, and the NE clump are well correlated spatially. We also notice that the region that located to the northeast of the black line is larger than the morphologically defined NE clump region (the red circle in Fig.~\ref{fig:galfit} and Fig.~\ref{fig:mse}). This is probably because the gas of the low-metallicity population is more extended than what can be identified by morphology.

\begin{figure*}
    \centering
    \includegraphics[width=1\textwidth,trim={0 0 0 0},clip]{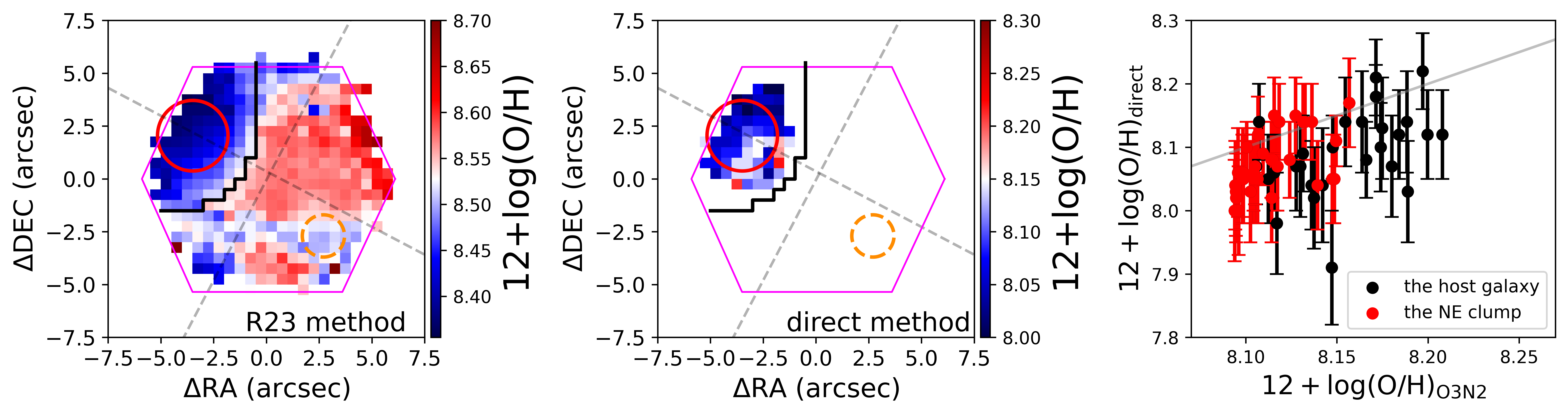}
     \caption{Left: The metallicity map obtained via the R23 method. Middle: The metallicity map obtained via the direct $T_e$ method. Only 61 spaxels with significant $\oiii\lambda$4363 detections ($\rm S/N>5$) are shown. Right: The comparison between the metallicities obtained from the direct method and those obtained from the O3N2 method for the 61 spaxels in the middle panel. The red data points and the black data points represent the spaxels inside and outside the morphologically defined NE clump (the red circle in the middle panel), respectively. The grey line is the one-to-one correlation.}
    \label{fig:2metallicity}
\end{figure*}

We mentioned earlier in this section that there are multiple methods to estimate the metallicity besides the O3N2 method. We check whether our main conclusions about the metallicity based on the O3N2 method still stand with the R23 method and the direct method.
In the left panel of Fig.~\ref{fig:2metallicity}, we show the metallicity map obtained using the R23 method \citep{tremonti2004}.
We can see that this metallicity map has the similar feature as the one obtained by the O3N2 method in the right panel of Fig.~\ref{fig:maps}. The metallicity of the NE clump is overall lower than that of the host galaxy.

In the MaNGA data cube, the $\oiii\lambda$4363 auroral line can be observed in some spaxels. This makes the direct method possible. From the MaNGA pipe3D \citep{pipe3da,pipe3db}, we select spaxels with the S/N of $\oiii\lambda4363$ larger than 5, and then visually check to see whether the $\oiii\lambda4363$ is indeed real. In the end, there are 61 spaxels selected to have the $\oiii\lambda$4363 detected. We calculate the metallicities of these 61 spaxels using the bayesisan model-based code {\tt HII-CHI-mistry~v5.1}  \citep{2014MNRAS.441.2663P} which uses the direct $T_e$ method. The spatial location and metallicity of the 61 spaxels are shown in the middle panel of Fig.~\ref{fig:2metallicity}. These spaxels are all in the low metallicity region defined in Fig.~\ref{fig:mse}, and show low metallicities.
In the right panel of Fig.~\ref{fig:2metallicity}, we compare the metallicity derived by the O3N2 method to that derived from the direct method. The metallicity given by the direct method is systematically lower ($\sim$ 0.1 dex) than that given by the O3N2 method.
Overall, Fig.~\ref{fig:2metallicity} shows that the metallicity of the NE clump is consistently lower than that of the host galaxy with various metallicity-measurement methods we checked.


\subsection{Kinematics}\label{sec:kinematic}

\begin{figure*}
    \centering
    \includegraphics[width=1\textwidth,trim=0 0 0 0]{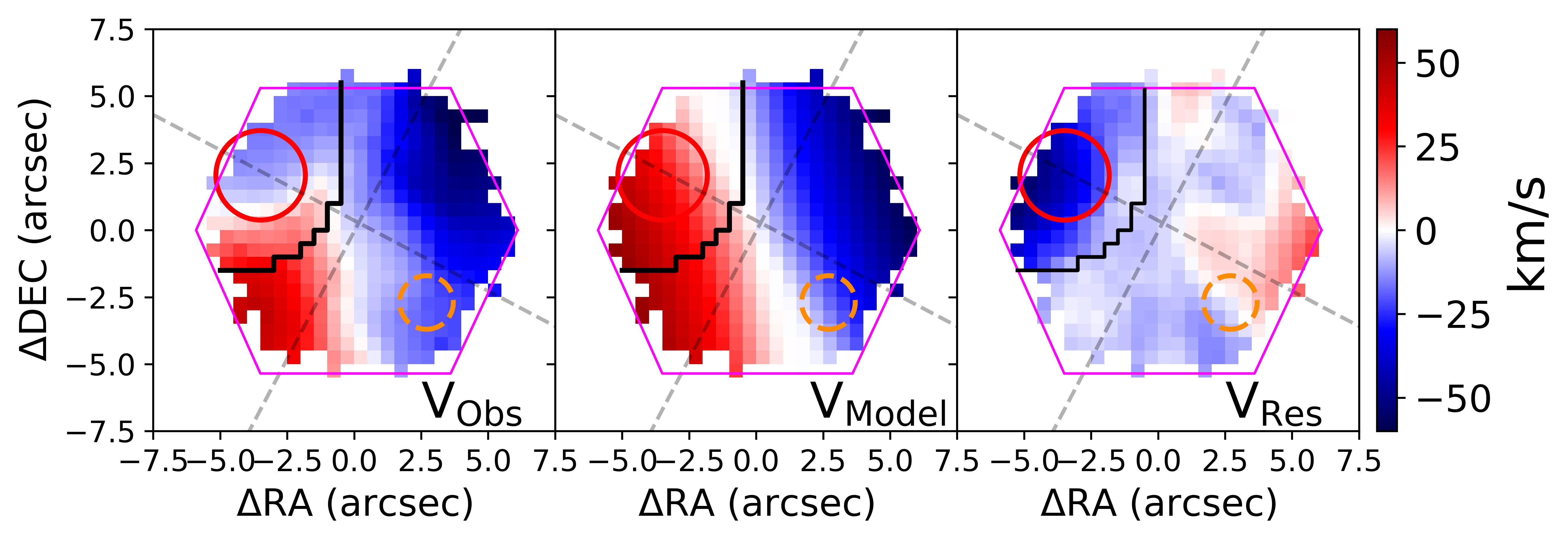}
    \caption{The observed H$\alpha$ velocity map (the left panel), the best-fit rotation model for the host galaxy (the middle panel), and the residuals ($V_{\text{obs}}-V_{\text{model}}$) of the H$\alpha$ velocity field (the right panel) of MaNGA8313-1901.}
    \label{fig:vel}
\end{figure*}

In the left panel of Fig.~\ref{fig:vel}, we show the H$\alpha$ velocity map taken from the MaNGA DAP data cube. The velocity field of the host galaxy is dominated by a rotating disk whose minor axis falls along the SE-NW direction. The NE clump lies close to the SW-NE major axis of the rotating disk of the host. The NE clump seems to be kinematically detached from the rotating disk of the host galaxy.
In the middle panel of Fig.~\ref{fig:vel}, we model the velocity field of the H$\alpha$ emission line of the host galaxy. The adopted rotation curve \citep{Andersen2013} is as follows:

\begin{equation}
	V(R) = v_{rot} \tanh(R/r_{rot}),
	\label{eq:3}
\end{equation}
where $r_{rot}$ is the turn-over radius, within which the rotation velocity increases with the radius $R$ until it reaches the maximum velocity $v_{rot}$ at $r_{rot}$. With Equation~(\ref{eq:3}), the 2D velocity field can then be obtained as

\begin{equation}
    V_{obs}(R,i) = V_\text{sys}+V(R)\cdot\sin i\cdot\cos \phi,
    \label{eq:4}
\end{equation}
where $V_\text{sys}$ is the systematic recession velocity, $V(R)$ is the intrinsic rotation curve as defined by Equation~(\ref{eq:3}), $i$ is the inclination angle of the rotating disk, and $\phi$ is the azimuthal angle in the galaxy plane. We use {\tt emcee} \citep{2013PASP..125..306F} to perform the velocity field fitting. In order to remove the contamination from the NE clump, we exclude in the fitting the metallicity defined NE clump region, i.e. the spaxels to the northeast of the black solid line (defined in Fig.~\ref{fig:mse}).

The best-fit velocity map is displayed in the middle panel of Fig.~\ref{fig:vel}, and the residual map is shown in the right panel. The parameters of the best-fit model are $i = 75.3^\circ \pm 14.4^\circ$, $\phi = -161.5^\circ \pm 9.8^\circ$, $v_\text{rot} = 90.7 \pm 32.5$ (km/s), and $r_\text{rot} =  6.9\arcsec\pm 1.6\arcsec$. The errors represent the $67\%$ confidence interval. From the residual map, we find that the NE clump has a systematic velocity towards us relative to the host galaxy, implying that this clump is a kinematically independent component. In this map, disturbance around the SW clump in the southwest corner is also weakly present.

Usually, bars in galaxies can also induce disturbances in the gas kinematics. However, the line-of-sight residual velocity of the gas at the NE clump region is quite significant at $\sim 25$km/s, almost comparable to the maximum rotation velocity of the gas in the MaNGA field. In addition, the maximum residual velocity tends to distribute toward the edge of the galaxy, instead of around the center. Thus we think the velocity disturbances of the gas at the NE clump is unlikely to be induced by a bar in the galaxy, which is weak at most, if it exists at all judging from its morphology.


\section{Discussion} \label{sec:discussion}

An enhanced star formation complex in a galaxy can be induced by galaxy interactions or an infall of gas accretion, or by secular processes, such as a bar. It can also be merely a giant star-bursting complex produced internally in the disk. Indeed, a similar case \citep{richards2014} was reported in the Sydney-Australian-Astronomical-Observatory Multi-object Integral-Field Spectrograph \citep[SAMI;][]{SAMI2012,SAMI2015} survey.  There is an off-centered luminous clump in a dwarf galaxy, GAMA J141103.98$\mbox{-}$003242.3, and the clump also shows lower metallicity and higher SFR compared to the host galaxy. However, in this galaxy, they found no obvious distortion in the kinematic map of the H$\alpha$ emission line, thus they proposed that this clump may just be an intrinsic stochasticity of star formation in a stable system of the host. This is different from MaNGA 8313-1901. In Fig.~\ref{fig:vel}, we show the H$\alpha$ velocity map and the residual velocity map after subtracting a rotating disk. The kinematics of the ionized gas in the NE clump suggests that it is detached from the host. This and the significant drop of the metallicity in the NE clump together point toward more to the scenario that the NE clump is triggered by external cause, such as galaxy interactions or gas accretion processes, rather than an internal star-bursting event. The much bigger size of the NE clump compared to local star-bursting complexes also hints its external origin. 

Therefore, in below we focus our discussions in external origin of the NE clump. We explore further the type of external origin the NE clump can have, mainly distinguishing between the gas accretion process and that induced by the interactions. Cold streams of pristine gas may directly flow into the disk and even to the central regions of a galaxy, particularly in a less-massive halo \citep{joung2012,peng2014}. According to \cite{Wang2016,WangH2018} catalog, MaNGA 8313-1901 seems to be in a large-scale filamentary structure, where the cold gas reservoir can be more abundant than in clusters. 
Its host halo mass is estimated to be M$_\text{halo} = 10^{11.03}\ \text{M}_{\sun}$ \citep{Yang2007,Yang2012}. In such a halo, the cold gas is more likely to fall into the galaxies compared to galaxies in massive halos \citep{Dekel2009}. Thus, the environment of MaNGA 8313-1901 can accommodate both external scenarios. 

In the gas accretion scenario, the accreted gas can come from the circumgalactic medium (CGM), the intergalactic medium (IGM), and/or even a companion galaxy, such as a gas-rich satellite \citep{Hwang2019}. The NE clump should contain only young stars that were just formed by the current star formation induced by the newly accreted gas. If the NE clump is induced by the galaxy-galaxy interactions instead, we would expect to see the old stellar populations from both the interacting galaxies. Thus we explore the origin of the NE clump mainly by checking whether there are any significant old stellar populations associated with it.

\begin{figure*}
    \centering
    \includegraphics[width=1\textwidth,trim=0 0 0 0]{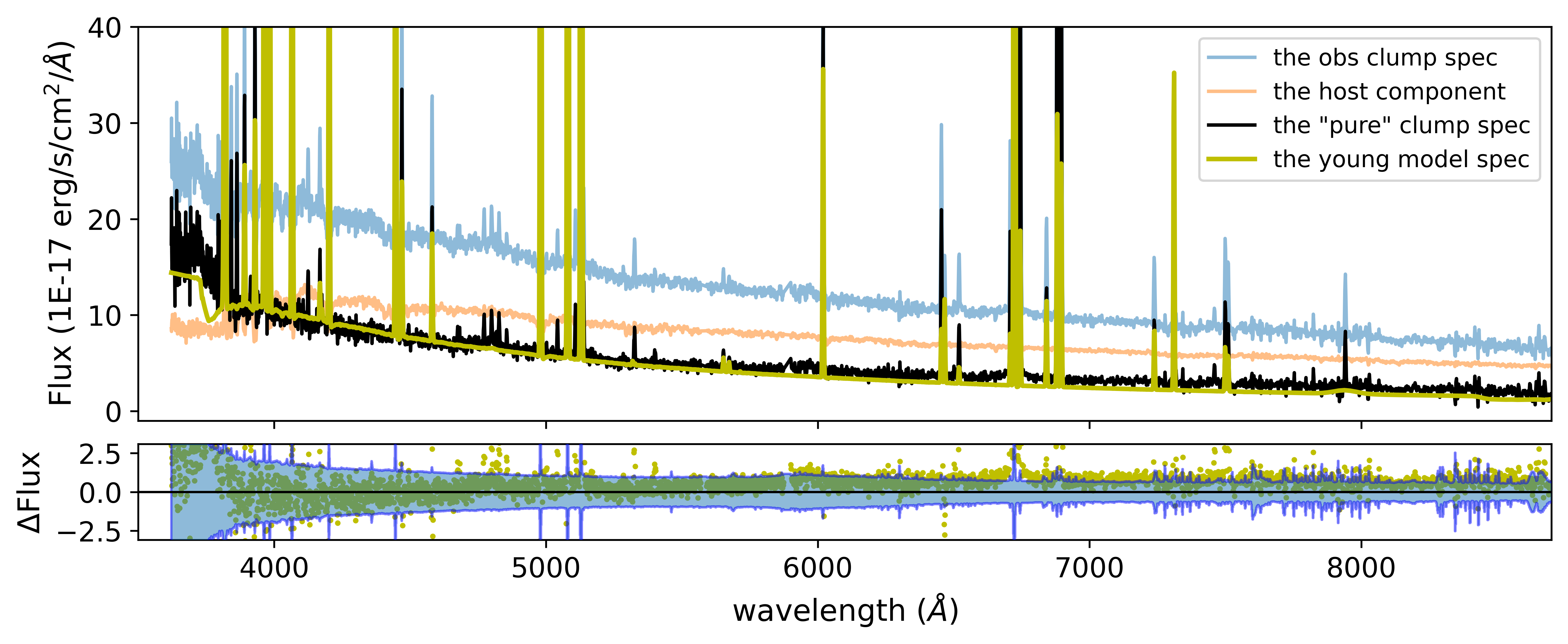}
    \caption{The spectra of different components in the NE clump of MaNGA 8313-1901. In the upper panel, The observed spectrum in the NE clump is shown in light blue. The light orange line represents the estimated host component in the NE clump. The ``pure" clump spectrum, i.e., the observed clump spectrum (light blue) subtracted by the estimated host spectrum (light orange), is shown in black. The yellow dashed spectrum shows the modeled clump spectrum  constructed by {\tt Prospector} (see text for details). The residual of the modeled ``pure" clump spectrum, i.e., the difference between the ``pure" clump spectrum and the modeled one, is shown in the lower panel. The blue shaded area shows the 3$\sigma$ error region of the observed spectrum.}
    \label{fig:sps}
\end{figure*}

The observed spectrum in the NE clump region, which we obtain by summing up the observed spectra of all spexals in the red circle, and shown as light blue in  Fig.~\ref{fig:sps}, includes light from both the NE clump and the underlying host galaxy in the same region. Therefore, to evaluate the stellar population of the NE clump itself, we need to subtract the  spectrum of the underlying host galaxy component in the NE region. This can be done by constructing a spectrum of the host galaxy component, by assuming a proper spectral shape, and then scaling it to the flux of the host galaxy component in the red circle.

\begin{figure}
    \centering
    \includegraphics[width=1.\textwidth,trim=0 0 0 0]{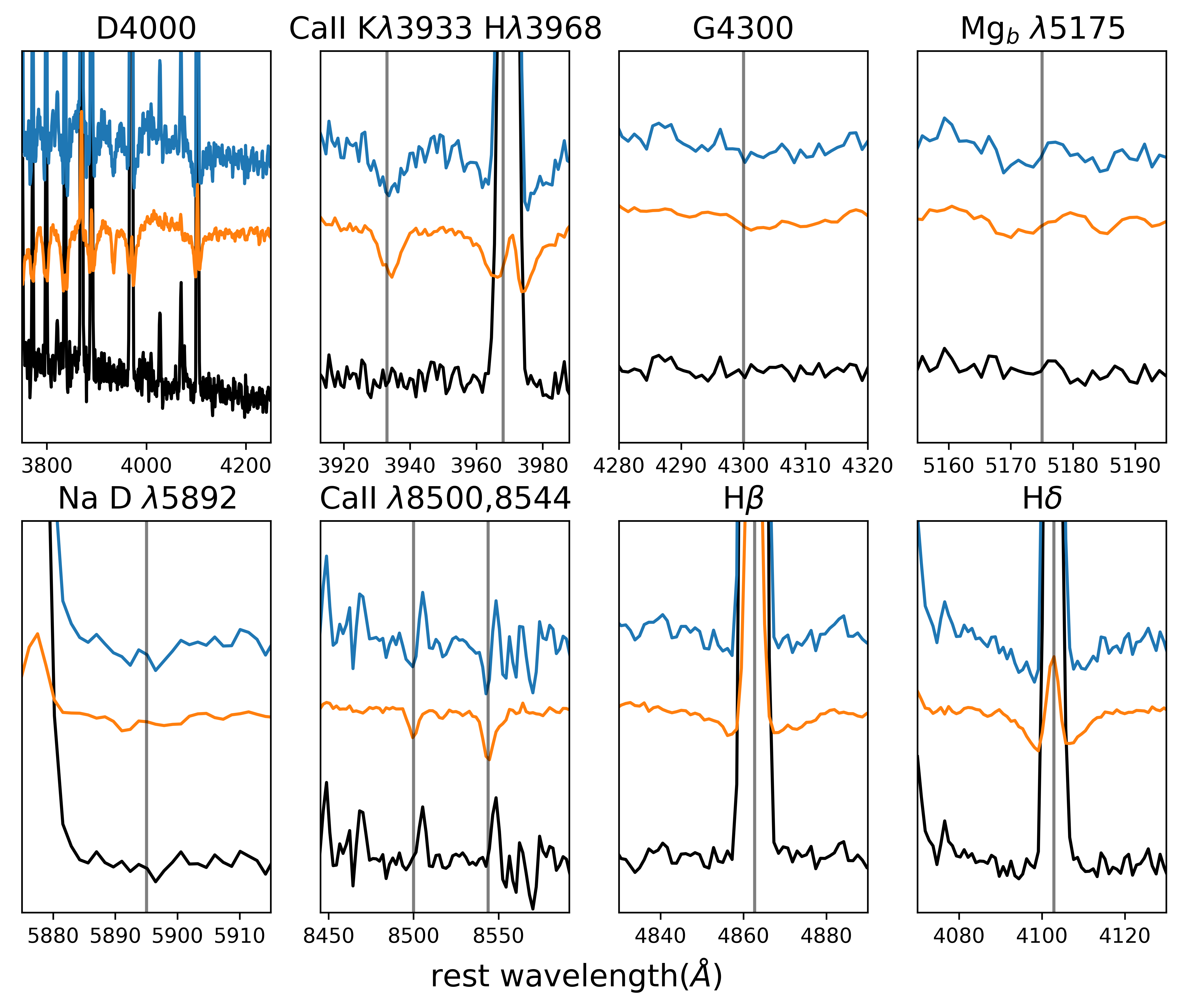}
    \caption{The zoom-in spectra of the observed spectrum (the blue line), the host component spectrum (the orange line), and the ``pure" clump spectrum (the black line). From top left to bottom right, we show the D4000 absorption lines, Ca\,{\scriptsize II} $\lambda3933,3968$, $G4300$, Mg$_b \lambda5175$, NaD $\lambda5892$ and  Ca\,{\scriptsize II} $\lambda8500,8544$, and emission lines H$\beta$ and H$\delta$.}
    \label{fig:lick}
\end{figure}

\begin{figure}
    \centering
    \includegraphics[width=1.\textwidth,trim=0 0 0 0]{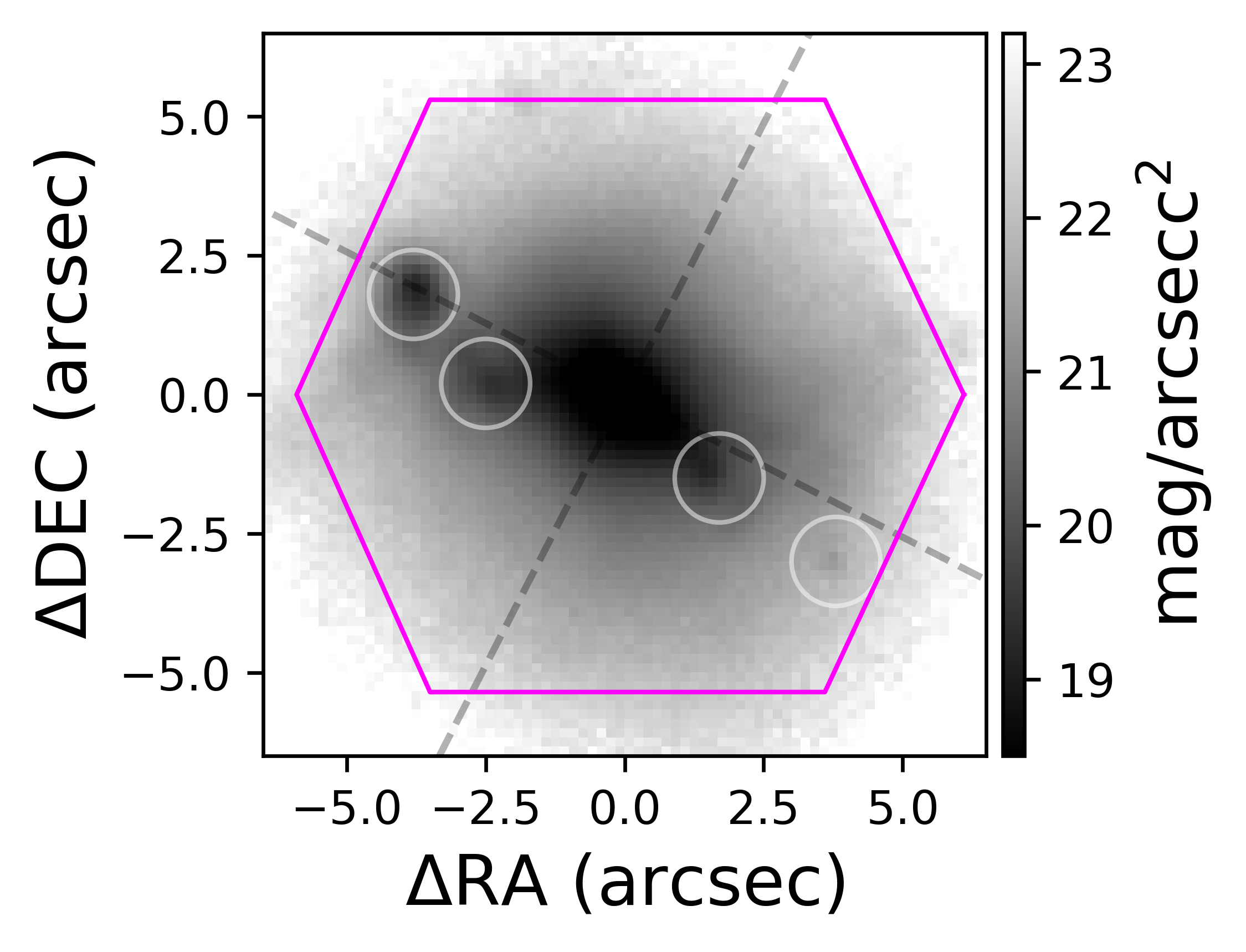}
    \caption{The $i$-band image of HSC survey. We draw four white circles to highlight the small knots that we find.}
    \label{fig:hsc}
\end{figure}

We use the image decomposition results obtained in Section~\ref{sec:DESI} to estimate the $g$-band flux of the underlying host component in the NE clump region in MaNGA 8313-1901. The structural parameters of the host galaxy obtained from the morphology decomposition (Fig.~\ref{fig:galfit}) is used to construct a full-scale (including the NE clump region) image of the pure host, which is then convolved with the PSF to produce what would be the observed $g$-band image of the pure host galaxy in MaNGA 8313-1901. The total $g$-band flux of the underlying host component within the NE clump region (the red circle in Fig.~\ref{fig:galfit}) can then be estimated from the constructed host image to be $\rm flux_{host\_in\_clump}$=10.83 $\times$ 10$^{-17}$ erg/s/cm$^2$. 

Next, we assume the spectral shape of the underlying host galaxy component in the NE region is the same as that at the galaxy center. This assumption is reasonable because the low mass galaxies often show flat slopes in their mass-to-light ratio (M/L) gradients, which suggests that the stellar population of the host galaxy is similar from the inside out \citep{gejunqiang2021}. We obtain a spectrum by summing up the spectra of all spexals within an aperture of a diameter of 2.5\arcsec around the galaxy center, and adopt its shape for the underlying host galaxy component. Together with the $g$-band flux of the underlying host component estimated above, we can then construct the spectrum of the underlying host component in the NE clump region. This is shown as the light orange spectrum in Fig.~\ref{fig:sps}.

Then we subtract the constructed underlying host-component spectrum (the light orange spectrum in Fig.~\ref{fig:sps}) from the observed spectrum (the light blue spectrum in Fig.~\ref{fig:sps}) to obtain the ``pure" clump spectrum (called so hearafterward), which is shown as the black line in Fig.~\ref{fig:sps}.  The strong continuum in the blue wavelengths and the weak Balmer break suggest that the stellar populations of the ``pure" NE clump are young \citep{Guseva2007}. 

In addition, the ``pure" clump spectrum is almost free of absorption lines. In Fig.~\ref{fig:lick}, we zoom in the observed NE clump spectrum (the blue line), the spectrum of the underlying host component in the NE clump region (the orange line), and the ``pure" clump spectrum (the black line) on some of the absorption lines often used to indicate stellar populations. Absorption lines, such as H$\beta$, H$\delta$, Ca\,{\scriptsize II} $\lambda3933,3968$, and Ca\,{\scriptsize II} $\lambda8500,8544$, are clearly seen in the observed spectrum (blue) and the underlying host spectrum (orange), but in the ``pure" clump spectrum (black), they all become very insignificant. In particular, a series of higher-ordered Balmer and $He$ absorption lines that appear in the first panel of Fig.~\ref{fig:lick} are all very well subtracted. Note that our orange spectrum is not a fit to the observed spectrum. It is constructed using just the spectral shape of the host galaxy and the flux obtained from the image analysis. Both information is independent from the observed spectrum in the NE clump region (the blue spectrum). Yet, the subtracted spectrum (black) of the blue and orange spectra show almost no absorption lines, indicating strongly that the ``pure" spectrum is indeed dominated by the young populations.

To give a quantitative evaluation on how young the ``pure" spectrum can be, we compare the obtained ``pure" spectrum with a synthesized spectrum of young stellar ages. We construct the spectrum using a software called {\tt Prospector}. {\tt Prospector} \citep{Johnson2021} accommodates both the single stellar population (SSP) models and the nebular emission models. It can produce a model spectrum with both a continuum and emission lines given a fix set of parameters, such as the age of this spectrum, the dust parameters, and the metallicity of the gas and stars. We list the details of our model as follows.

\begin{itemize}
\item SSP models are from the {\tt FSPS} software \citep{conroy_propagation_2009,conroy_propagation_2010} which is connected to {\tt Prospector} by python-fsps \citep{foreman-mackey_python-fsps_2014}. The stellar spectral library, isochrone model and stellar initial mass function (IMF) we used are {\tt MILES} \citep{2006MNRAS.371..703S}, {\tt MIST} \citep{choi2016} and the Salpeter IMF \citep{Salpeter1955}, respectively. To model the spectrum of a metal-poor component, we fix the metallicity [Z/H] at $-0.4$ according to the median value of the metallicity map in the host galaxy region (the right panel of Fig.~\ref{fig:maps}).

\item The {\tt CLOUDY} software \citep{Ferland2013,Byler2017} is used to model the nebular emission. The metallicity of gas, [Z/H]$_{\text{gas}}$, is set to be $-0.6$ according to the median value of the metallicity map in the NE clump region (the right panel of Fig.~\ref{fig:maps}). 

\item The dust attenuation model we used \citep{Kriek2013} consists of two components: the dust in the birth cloud around young stars and the diffuse dust in the interstellar medium. The optical depth of the diffuse dust, $\tau_{\text{diff}}$, is 0.001. There is a degeneracy between the stellar population and the dust attenuation -- a younger stellar population is expected when using a higher dust attenuation. We use the default value of the optical depth ratio of the birth cloud dust to the diffuse dust, $\tau_{\text{bc}}/\tau_{\text{diff}} = 1$, and the default value of the diffuse dust index, $n_{\text{diff}} = 0$, which regulates the shape of the attenuation curve.
\end{itemize}

With these parameters set, we use {\tt Prospector} to generate a spectrum with just $\le$ 7 Myr stellar population, and show the result in yellow in Fig.~\ref{fig:sps}. The model has a continuous constant SFR within the last $0-7$ Myr with a total stellar mass of 1.8 $\times 10^6$ M$_{\odot}$. The synthesized model spectrum matches very well with the black spectrum. The residual between the model and the ``pure'' clump spectrum is very small, and we show it in the lower panel in Fig.~\ref{fig:sps}. Also shown in the lower panel is the blue shaded region, which is the 3$\sigma$ region of the observed spectrum of the NE clump (the light blue spectrum). The errors are estimated from the observed flux errors of the spectra of all the spexals in the red circle, also considering the covariance of the spexals \citep{law2016}. Most of the residuals are within the errorbars of the observed spectrum.

The extremely young stellar population required by the good-matching synthesized model seems to suggest that the dominating stellar populations in the ``pure" NE clump are very young and the old stellar populations are quite insignificant. Based on the argument we listed earlier, this may imply that the NE clump is  triggered by the gas accretion scenario.

Of course, this analysis cannot exclude scenarios of mergers of gas-rich dwarfs that do not have significant old stellar populations compared to the host galaxy of MaNGA 8313-1901 from the beginning. Even though insignificant, we try to give a quantitative constraint on the amount of the old stellar population the ``pure" NE clump can have. We notice that within the errorbars, the residual spectrum of the ``pure" NE clump subtracted by the 7-Myr young stellar population model spectrum does show offset that seems to grow increasingly bigger at longer wavelengths. If this offset is real, it could be accounted for by an old stellar population component. We use the difference of the $z$-band flux between the model spectrum (estimated to be 1.57$\times$ 10$^{-17}$ erg/s/cm$^2$) and the ``pure" clump spectrum (estimated to be 1.98 $\times$ 10$^{-17}$ erg/s/cm$^2$) to estimate the corresponding mass of this possible old stellar population. We choose an SSP template of 10 Gyr old and [Z/H]$_{\text{star}}=0$ to make the estimation, the age of the SSP is chosen to be old enough so that the mass is unlikely to be underestimated. Taken a M/L ratio at $z$-band of 1.72 (M$_\sun/L_{\sun,z}$) for such a stellar population \citep{bc03}, we estimate the mass of this possible old SSP to be about 3.89 $\times 10^7$ M$_{\odot}$. This is only $\sim 20\%$ of the mass of the underlying host component in the red circle, and $\sim 1\%$ of the mass of the total host galaxy. Remember that the ``pure" clump spectrum (the black spectrum in Fig.~\ref{fig:sps}) is obtained via a very simple subtraction method, there can have a lot of uncertainties that can easily account for the $\sim 20\%$ differences in the flux. Our synthesized model spectrum (the yellow spectrum in Fig.~\ref{fig:sps}) is also simple. Thus, we do not make an attempt to interpret these evaluations further. We think it is safe to say that even if the old stellar population exists, it is very insignificant compared to the host galaxy.

The mass of the \hi~gas of MaNGA 8313-1901 is log(M$_{\text{HI}}$/M${_\odot}$) = 9.37 \citep{HI2019}.
Combining with its stellar mass log(M$_{\text{*}}$/M${_\odot}$) = 8.88 (also listed in Table~\ref{tab:table}), 
the gas-to-stellar mass ratio of MaNGA 8313-1901 is about 3, which is quite typical for galaxies with similar stellar masses. This suggests that there is no obvious gas excess in this galaxy. In the gas accretion scenario, since the metallicity of the NE clump is $\sim 0.1-0.2$ dex lower than that of the host galaxy (Fig.~ \ref{fig:mse} and Fig.~\ref{fig:2metallicity}), we speculate that the accreted gas is metal-poor, otherwise it should require a significant amount of accreting gas to lower the metallicity of a big index as observed in our case. Assuming that the accreted gas has a metallicity one tenth of that of the host ($\rm Z_{gas}=Z_{host}/10$), to lower the metallicity in the NE clump region by 0.15 dex, the mass of the accreted gas should be about half of that of the gas already existed in the NE clump region. Considering that there is no obvious gas excess observed in this galaxy, the metallicity of the accreted gas would then be expected to be significantly lower than that of the host, possibly even be the pristine gas. 


In addition, the NE clump is located on the major axis of MaNGA 8313-1901. This probably means that the gas is accreted along this direction. Due to the projection effect, we are not certain if the accretion direction is along the disk plane, or have an angle with it. However, it is interesting to see that the SW clump (the orange circle in all the 2D maps shown in this paper) mentioned in Section~\ref{sec:sfr} is also found on the major axis. Even though the metallicity (the right panel of Fig.~\ref{fig:maps}), the EW(H$\alpha$) (the bottom right panel of Fig.~\ref{fig:image}) and the kinematics (Fig.~\ref{fig:vel}) of the SW clump are all only marginally different from its surroundings, we still think that the SW clump is similar to the NE clump and has an external origin, instead of being a part of the disk. A suggestive clue comes from Fig.~\ref{fig:hsc}, where we show the $i$-band image from the Hyper Suprime-Cam Subaru Strategic Program on the 8.2 m Subaru Telescope \citep[HSC-SSP;][]{HSC,HSCdr3}. The image is deeper and is of higher spatial resolution compared to the SDSS or the DESI Legacy images. The image shows that there may exist multiple clumps including the NE and the SW clumps, distributed along the major axis of the galaxy, which we tentatively mark with white circles in Fig.~\ref{fig:hsc}. They may all be part of the same accretion event. Images with even higher spatial resolution are needed to explore this further.

\section{Summary}\label{summary}

In this paper, we report a BCD galaxy, MaNGA 8313-1901, that has a distinct off-centered clump to the northeast direction which we name as the NE clump. We use {\tt GALFIT} to obtain the structural information of the host galaxy and the NE clump. We fit the host galaxy and the NE clump with a S\'ersic model convolving the PSF. The effective radius and the S\'ersic index of the host galaxy are $r_e=1.23\pm0.13$ kpc and $n=1.46\pm0.04$. Those of the NE clump are $r_e=290\pm12$ pc and $n=0.19\pm0.04$. The size of the NE clump is significantly larger than typical \hii~regions seen in local galaxies, and is more similar to the starburst clumps in the high-$z$ galaxies. 

We derive the SFR map and the gas-phase metallicity map for this galaxy from the MaNGA data cube. The NE clump shows features of enhanced star-formation and lowered metallicity. The result that the NE clump has lower metallicity compared to the host stays with different metallicity indicators. We further study the gas kinematic map as inferred from the H$\alpha$ emission lines. By building a kinematic model of the H$\alpha$ velocity map, we find that the NE clump is kinematically detached from the rotating disk of the host galaxy. The lowered metallicity, enhanced star formation, and detached kinematics of the NE clump all imply that the clump is not internally originated as part of the disk, such as an \hii~complex. Instead, we prefer the external origin for the NE clump. Processes such as the gas accretion or galaxy mergers can all produce big clumps with characters similar to the NE clump.
We further analyze the stellar populations in the NE clump to explore whether the NE clump contains old stars of its own, aiming to distinguish the scenarios between the gas accretion and the galaxy interaction with significant pre-existing old stars. 

In order to study the stellar population of the NE clump itself, we need to properly subtract the underlying host component from the observed spectrum of the NE clump region. We assume that the spectral shape of the host component in the NE clump is the same as the center of the host galaxy. We scale the host spectra with the flux of the host component in the NE clump estimated from our best-fit morphological S\'ersic model. This spectra of the underlying host component in the NE clump region is shown in Fig.~\ref{fig:sps} as light orange. The subtraction of the observed clump spectrum (light blue) and the underlying host spectrum (light orange) results in the ``pure" clump spectrum (black in Fig.~\ref{fig:sps}). It shows obvious characteristics of young stellar populations such as the blue continuum and is almost free of absorption lines. 

We use the {\tt Prospector} to construct a spectrum with only young stellar populations ($\le7$ Myr) combined with dust, nebular continuum and emission lines, which is shown as the yellow spectrum in Fig.~\ref{fig:sps}. The modeled spectrum agrees very well with the ``pure" clump spectrum (black in Fig.~\ref{fig:sps}). This suggests that the ``pure" clump spectrum indeed predominantly consists of young stellar populations. The lack of having significant old stars in the NE clump favors the gas accretion scenario that the gas in the NE clump may be newly accreted. Of course, we can not exclude scenarios of mergers of gas-rich dwarfs that do not have significant old stellar population compared to the host galaxy of MaNGA 8313-1901 from the beginning.  

Besides the NE clump, there is another low-metallicity region to the southwest of the host galaxy, and we call it the SW clump. The SW clump shows similar characteristics to the NE clump in the SFR, the gas-phase metallicity, and the kinematics, but to a less extent compared with the NE clump. Both the two clumps are located on the major axis of the host galaxy. We speculate that the NE and SW clumps might have similar external origins. 

The results of this study can help us better understand the gas accretion in high-$z$ galaxies. Future moderately-deep infrared observations can certainly increase our diagnosing power on the old-stellar substrate in the NE clump. Further studies with the \hi~gas content of the galaxy, such as the observations with the Five-hundred-meter Aperture Spherical radio Telescope (FAST) and future resolved \hi~ measurements from the Square Kilometre Array (SKA), can be very critical. In addition, the future IFU observation in the optical with the Chinese Space Station Telescope (CSST-IFU), can also help to further dissect the clump. The more detailed investigation of the host properties of MaNGA 8313-1901, combined with more BCDs, can also tell us more of their evolution scenarios. In fact, we have compiled a sample of BCDs from the MaNGA survey and found at least 2 other BCDs that may host off-centered clumps, even though the clumps are not as significant as shown in MaNGA 8313-1901. We will study them in more detail in the future. 


\section*{Acknowledgements}
We thank the anonymous referee for his/her constructive comments that significantly helped the improvements of the manuscript. This work is supported by the National Key R\&D Program of China (No. 2019YFA0405501), the National Natural Science Foundation of China (No. U1831205, 12073059, \& U2031139).
J.Y. acknowledges support by the Natural Science Foundation of Shanghai (Project Number: 22ZR1473000). We also acknowledge the science research grants from the China Manned Space Project with NO. CMS-CSST-2021-A07, CMS-CSST-2021-A08, CMS-CSST2021-A09, CMS-CSST-2021-B04.
S.F. is supported by National Natural Science Foundation of China (No. 12103017), Natural Science Foundation of Hebei Province (No. A2021205001) and Science Foundation of Hebei Normal University (No. L2021B08).
R.R. thanks to Conselho Nacional de Desenvolvimento Cient\'{i}fico e Tecnol\'ogico ( CNPq, Proj. 311223/2020-6,  304927/2017-1 and  400352/2016-8), Funda\c{c}\~ao de amparo \'a pesquisa do Rio Grande do Sul (FAPERGS, Proj. 16/2551-0000251-7 and 19/1750-2), Coordena\c{c}\~ao de Aperfei\c{c}oamento de Pessoal de N\'{i}vel Superior (CAPES, Proj. 0001). 
J.G.F-T gratefully acknowledges the grant support provided by Proyecto Fondecyt Iniciaci\'on No. 11220340, and also from ANID Concurso de Fomento a la Vinculaci\'on Internacional para Instituciones de Investigaci\'on Regionales (Modalidad corta duraci\'on) Proyecto No. FOVI210020, and from the Joint Committee ESO-Government of Chile 2021 (ORP 023/2021). 

Funding for the Sloan Digital Sky Survey IV has been provided by the Alfred P.Sloan Foundation, the U.S. Department of Energy Office of Science, and the Participating Institutions. SDSS-IV acknowledges support and resources from the Center for High-Performance Computing at the University of Utah. The SDSS web site is www.sdss.org.

SDSS-IV is managed by the Astrophysical Research Consortium for the Participating Institutions of the SDSS Collaboration including the Brazilian Participation Group, the Carnegie Institution for Science, Carnegie Mellon University, Center for Astrophysics — Harvard \& Smithsonian, the Chilean Participation Group, the French Participation Group, Instituto de Astrof\'isica de Canarias, The Johns Hopkins University, Kavli Institute for the Physics and Mathematics of the Universe (IPMU) / University of Tokyo, the Korean Participation Group, Lawrence Berkeley National Laboratory, Leibniz Institut f\"ur Astrophysik Potsdam (AIP), Max-Planck-Institut f\"ur Astronomie (MPIA Heidelberg), Max-Planck-Institut f\"ur Astrophysik (MPA Garching), Max-Planck-Institut f\"ur Extraterrestrische Physik (MPE), National Astronomical Observatories of China, New Mexico State University, New York University, University of Notre Dame, Observat\'ario Nacional / MCTI, The Ohio State University, Pennsylvania State University, Shanghai Astronomical Observatory, United Kingdom Participation Group, Universidad Nacional Aut\'onoma de M\'exico, University of Arizona, University of Colorado Boulder, University of Oxford, University of Portsmouth, University of Utah, University of Virginia, University of Washington, University of Wisconsin, Vanderbilt University, and Yale University.

\software{GALFIT \citep[v3.0.5;][]{peng2002,peng2010}, MEGACUBE \citep{Mallmann2018,Riffel2021}, Prospector \citep{Johnson2021}, FSPS \citep{conroy_propagation_2009,conroy_propagation_2010}, python-fsps \citep{foreman-mackey_python-fsps_2014}, CLOUDY software \citep{Ferland2013,Byler2017}}

\bibliographystyle{aasjournal}
\bibliography{cita.bib}

\end{CJK*}
\end{document}